%% file: main.tex
\documentclass[runningheads]{llncs}

\usepackage{booktabs} 
\usepackage{bibnames}

\newcommand{\repeatthanks}{\textsuperscript{\thefootnote}}

\usepackage[ruled, vlined, linesnumbered, noend]{algorithm2e}
\usepackage{amsmath,bm,mathtools}
\usepackage{amssymb}
\usepackage{graphicx}
\usepackage{subfig}
\usepackage{xcolor}
\usepackage{multirow}
\usepackage{hyperref}

\usepackage{listings}
\usepackage{xcolor}
\usepackage{svg}
\usepackage[firstpage]{draftwatermark}
\SetWatermarkText{\hspace*{5in}\raisebox{7.5in}{\includegraphics[scale=0.1]{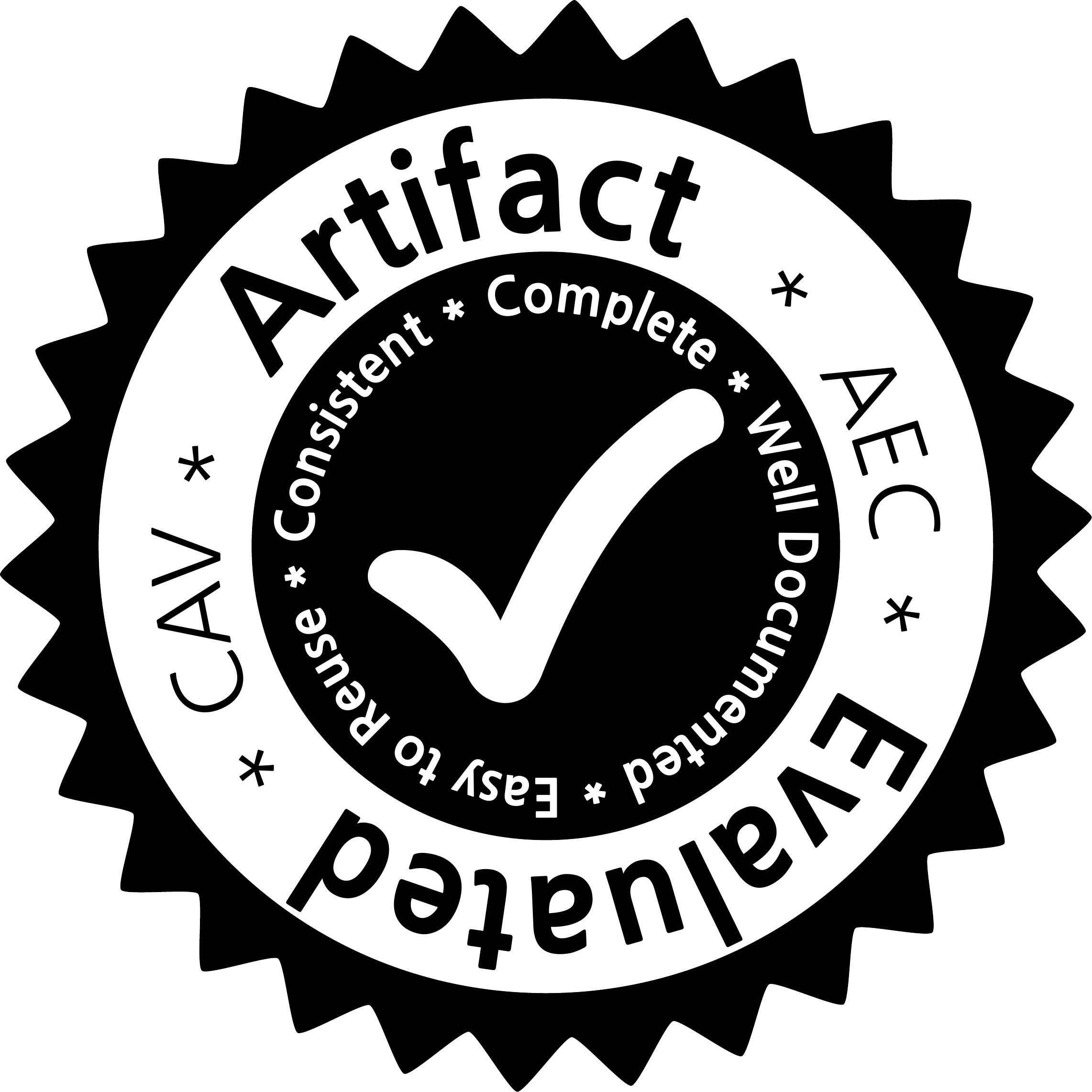}}}
\SetWatermarkAngle{0}
 
\definecolor{codegreen}{rgb}{0,0.6,0}
\definecolor{codegray}{rgb}{0.5,0.5,0.5}
\definecolor{codepurple}{rgb}{0.58,0,0.82}
\definecolor{backcolour}{rgb}{0.95,0.95,0.92}
 
\lstdefinestyle{mystyle}{
    backgroundcolor=\color{backcolour},   
    commentstyle=\color{codegreen},
    keywordstyle=\color{magenta},
    numberstyle=\tiny\color{codegray},
    stringstyle=\color{codepurple},
    basicstyle=\ttfamily\scriptsize,
    breakatwhitespace=false,         
    breaklines=true,                 
    captionpos=b,                    
    keepspaces=true,                 
    numbers=left,                    
    numbersep=5pt,                  
    showspaces=false,                
    showstringspaces=false,
    showtabs=false,                  
    tabsize=2
}
 
\allowdisplaybreaks

\newcommand{\minitab}[2][l]{\begin{tabular}{#1}#2\end{tabular}} 

\SetCommentSty{mycommfont}

\begin{document}
\title{SAW: A Tool for Safety Analysis of \\ Weakly-hard Systems\thanks{This work is supported by the National Science Foundation awards 1834701, 1834324, 1839511, 1724341, and the Office of Naval Research grant N00014-19-1-2496. It is also supported by the Asian Office of Aerospace Research and Development (AOARD), jointly with the Office of Naval Research Global (ONRG), award FA2386-19-1-4037, the Taiwan Ministry of Education (MOE) grants NTU-107V0901 and NTU-108V0901, the Taiwan Ministry of Science and Technology (MOST) grants MOST-108-2636-E-002-011 and MOST-109-2636-E-002-022.}}

\author{Chao Huang\inst{1}\thanks{Chao Huang and Kai-Chieh Chang contributed equally.} \and Kai-Chieh Chang \inst{2}\repeatthanks \and Chung-Wei Lin\inst{2} \and Qi Zhu\inst{1}}

\authorrunning{Chao et al.}

\institute{Northwestern University, \email{\{chao.huang,qzhu\}@northwestern.edu} \and National Taiwan University, \email{551100kk@gmail.com}, \email{cwlin@csie.ntu.edu.tw}}

\maketitle

\begin{abstract}
    We introduce SAW, a tool for safety analysis of weakly-hard systems, in which traditional hard timing constraints are relaxed to allow bounded deadline misses for improving design flexibility and runtime resiliency. Safety verification is a key issue for weakly-hard systems, as it ensures system safety under allowed deadline misses. Previous works are either for linear systems only, or limited to a certain type of nonlinear systems (e.g., systems that satisfy exponential stability and Lipschitz continuity of the system dynamics). In this work, we propose a new technique for infinite-time safety verification of general nonlinear weakly-hard systems. Our approach first discretizes the safe state set into grids and constructs a directed graph, where nodes represent the grids and edges represent the reachability relation. Based on graph theory and dynamic programming, our approach can effectively find the safe initial set (consisting of a set of grids), from which the system can be proven safe under given weakly-hard constraints. Experimental results demonstrate the effectiveness of our approach, when compared with the state-of-the-art.
    An open source implementation of our tool is available at \url{https://github.com/551100kk/SAW}. The virtual machine where the tool is ready to run can be found at \url{https://www.csie.ntu.edu.tw/~r08922054/SAW.ova}.
\end{abstract}

\input{sections/intro}

\input{sections/approach}

\input{sections/usage}

\input{sections/experiment}

\input{sections/conclusion}

\clearpage

\bibliographystyle{splncs03}
\bibliography{chao,hyoseung,references_qi,wenchao,bib_zwang}


\end{document}


\title{ReachNN: Reachability Analysis of Neural-Network\\ Controlled Systems -- Appendix}

\maketitle

\input{sections/appendix.tex}


%% file: sections/intro.tex
\section{Introduction} \label{sec:intro}

Hard timing constraints, where deadlines should always been met, have been widely used in real-time systems to ensure system safety. However, with the rapid increase of system functional and architectural complexity, hard deadlines have become increasingly pessimistic and often lead to infeasible designs or over provisioning of system resources~\cite{Zhu_PIEEE18,Lin_TODAES15,Huang_DESTION19,Liang_ICCD19}. 
The concept of weakly-hard systems are thus proposed to relax hard timing constraints by allowing occasional deadline misses~\cite{Hamdaoui_IEEE95,Bernat_TC01}. This is motivated by the fact that many system functions, such as some control tasks, have certain degrees of robustness and can in fact tolerate some deadline misses, as long as those misses are bounded and dependably controlled. In recent years, considerable efforts have been made in the research of weakly-hard systems, including schedulability analysis~\cite{Bernat_TC01,Li_IEEE06,Sun_TECS17,Quinton_DATE_12,Hammadeh_EMSOFT_14,Xu_ECRTS_15,Hammadeh_DATE17,Hammadeh_ECRTS_17,Ahrendts_ECRTS18,Choi_RTAS19}, opportunistic control for energy saving \cite{huang2020opportunistic}, control stability analysis and optimization~\cite{Frehse_RTSS_14,Ramanathan_TPDS_99,Gaid_IFAC_08,Marti_II_10,Pazzaglia_ECRTS18}, and control-schedule co-design under possible deadline misses~\cite{Damoon_IEEE18,Tobias_14,Chwa_RTAS18}. Compared with hard deadlines, weakly-hard constraints can more accurately capture the timing requirements of those system functions that tolerate deadline misses, and significantly improve system feasibility and flexibility~\cite{Liang_ICCD19,Huang_DESTION19}. Compared with soft deadlines, where any deadline miss is allowed, weakly-hard constraints could still provide deterministic guarantees on system safety, stability, performance, and other properties under formal analysis~\cite{Huang_HSCC19,wardega-date20}.

\begin{figure}[tbp]
	\centering	
	\includegraphics[width=0.8\textwidth]{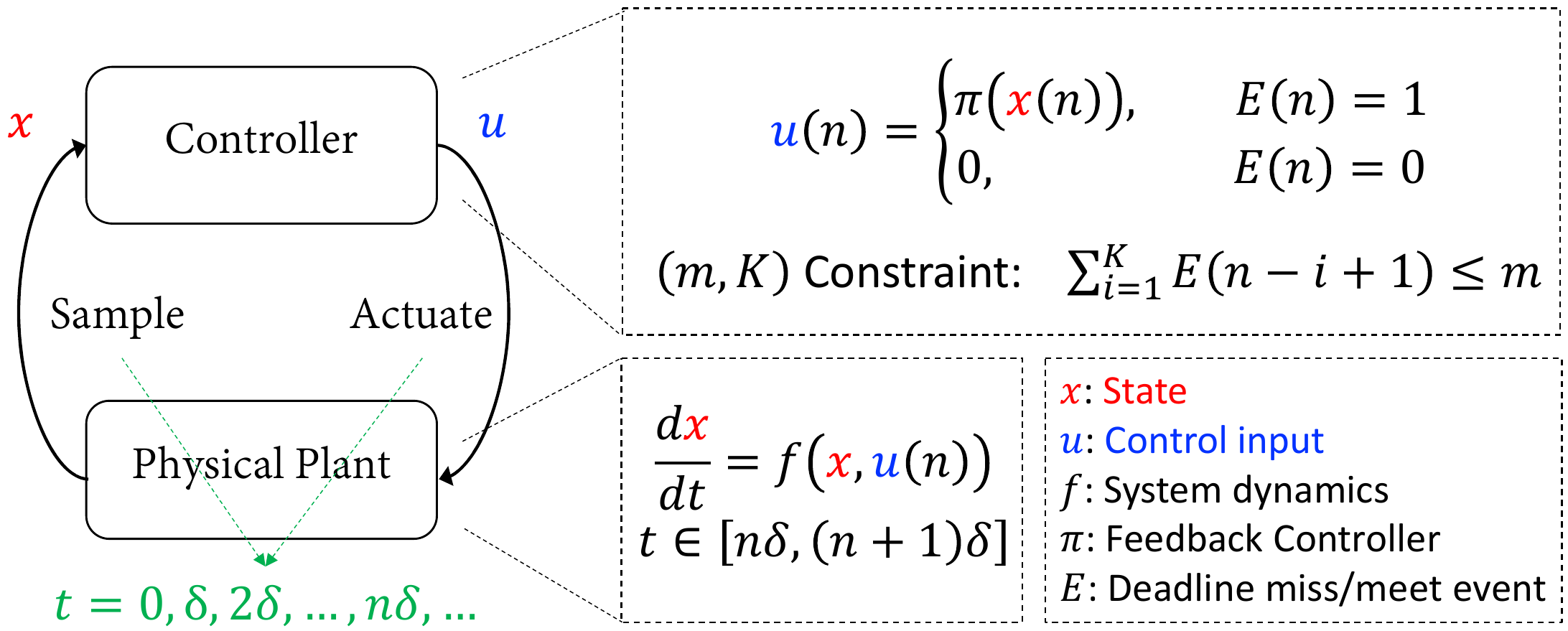}%
	\caption{A weakly-hard system with perfect sensors and actuators.}
	\label{fig:whs}
\end{figure}

A common type of weakly-hard model is the $(m, K)$ constraint, which specifies that among any $K$ consecutive task executions, at most $m$ instances could violate their deadlines~\cite{Bernat_TC01}. Specifically, the high-level structure of a $(m,K)$-constrained weakly-hard system is presented in Figure~\ref{fig:whs}. Given a sampled-data system $\dot{x} = f(x,u)$ with a sampling period $\delta > 0$, the system samples the state $x$ at the time $t = i\delta$ for $n=0,1,2,\dots$, and computes the control input $u$ with function $\pi(x)$. 
If the computation completes within the given deadline, the system applies $u$ to influence the plant's dynamics. Otherwise, the system stops the computation and applies zero control input. As aforementioned, the system should ensure the control input can be successfully computed and applied within the deadline for at least $K{-}m$ times over any $K$ consecutive sampling periods.

For such weakly-hard systems, a natural and critical question is whether the system is safe by allowing deadline misses defined in a given $(m,K)$ constraint. There is only limited prior work in this area, while nominal systems have been adequately studied \cite{huang2017probabilistic,yang2016linear,frehse2011spaceex,chen2013flow}.
In~\cite{Frehse_RTSS_14}, a weakly-hard system with linear dynamic is modeled as a hybrid automaton and then the reachability of the generated hybrid automaton is verified by the tool SpaceEx~\cite{frehse2011spaceex}. In~\cite{duggirala2015analyzing}, the behavior of a linear weakly-hard system is transformed into a program, and program verification techniques such as abstract interpretation and SMT solvers can be applied. 

In our previous work~\cite{Huang_HSCC19}, the safety of nonlinear weakly-hard systems are considered for the first time. Our approach tries to derive a safe initial set for any given $(m,K)$ constraint, that is, starting from any initial state within such set, the system will always stay within the same safe state set under the given weakly-hard constraint. Specifically, we first convert the infinite-time safety problem into a finite one by finding a set satisfying both \emph{local safety} and \emph{inductiveness}. The computation of such valid set heavily lies on the estimation of the system state evolution, where two key assumptions are made: 1) The system is exponentially stable under nominal cases without any deadline misses, which makes the system state contract with a constant decay rate; 2) The system dynamics are Lipschitz continuous, which helps bound the expansion under a deadline miss. Based on these two assumptions, we can abstract the safety verification problem as a one-dimensional problem and use linear programming (LP) to solve it, which we call \emph{one-dimension abstraction} in the rest of the paper. 

In practice, however, the assumptions in~\cite{Huang_HSCC19} are often hard to satisfy and the parameters of exponential stability are difficult to obtain. In addition, while the scalar abstraction provides high efficiency, the experiments demonstrate that the estimation is always over conservative.
In this paper, we go one step further and present a new tool SAW for infinite-time safety verification of nonlinear weakly-hard systems \textbf{without any particular assumption on exponential stability and Lipschitz bound}, and try to be less conservative than the scalar abstraction. Formally, the problem solved by this tool is described as follows:

\begin{problem}
    Given an $(m,K)$ weakly-hard system with nonlinear dynamics $\dot{x}=f(x,u)$, sampling period $\delta$, and safe set $X$, find a safe initial set $X_0$, such that from any state $x(0) \in X_0$, the system will always be inside $X$.  
\end{problem}

To solve this problem, we first discretize the safe state set $X$ into grids. We then try to find the grid set that satisfies both local safety and inductiveness. For each property, we build a directed graph, where each node corresponds to a grid and each directed edge represents the mapping between grids with respect to reachability. We will then be able to leverage graph theory to construct the initial safe set. Experimental results demonstrate that our tool is effective for general nonlinear systems.

%% file: sections/approach.tex
\section{Algorithms and Tool Design} \label{sec:approach}

\begin{figure}[t]
\centering
\includegraphics[width=1\textwidth]{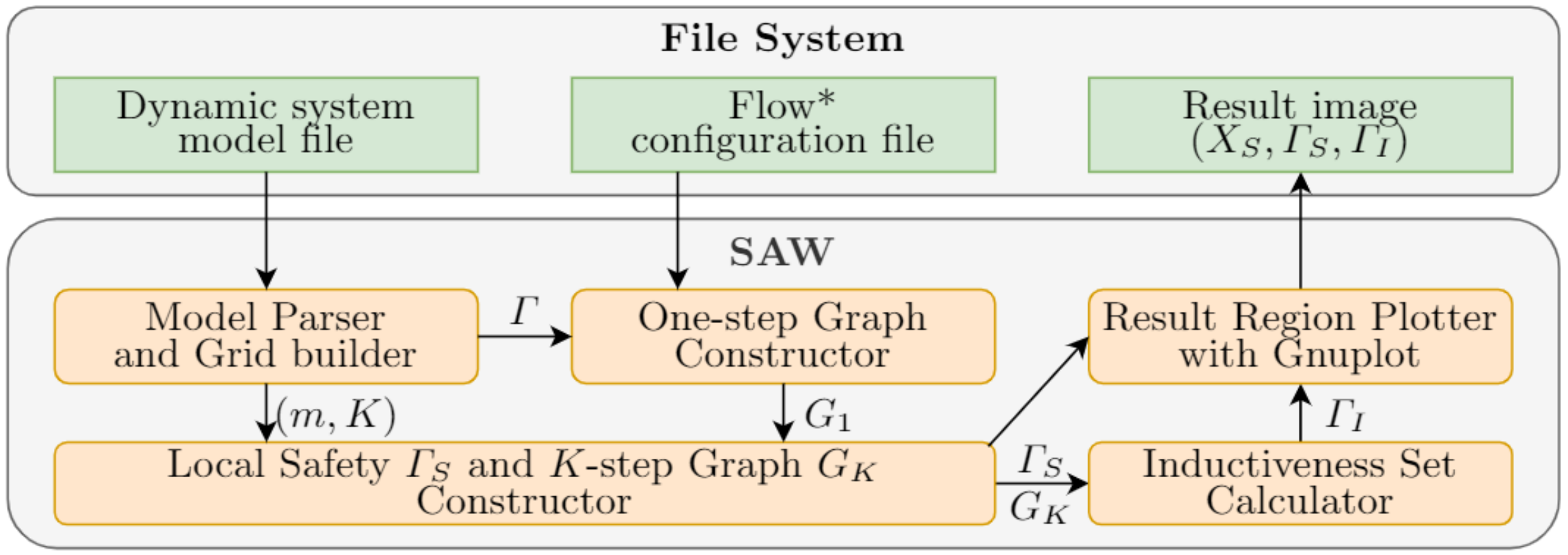}
\caption{The schematic diagram of SAW.}
\label{fig:schematic}
\end{figure}

The schematic diagram of our tool SAW is shown in Figure~\ref{fig:schematic}. The input is a model file that specifies the system dynamics, sampling period, safe region and other parameters, and a configuration file of Flow*~\cite{chen2013flow} (which is set by default but can also be customized). After fed with the input, the tool works as follows (shown in Algorithm~\ref{algo:safe}). The safe state set $X$ is first uniformly partitioned into small grids $\Gamma = \{v_1,v_2,\ldots,v_{p^d}\}$, where $X = v_1 \cup v_2 \cup \cdots \cup v_{d^p}$, $v_i \cap v_j = \phi$ ($\forall i\neq j$), $d$ is the dimension of the state space, and $p$ is the number of partitions in each dimension (Line 1 in Algorithm~\ref{algo:safe}). The tool then tries to find the grids that satisfy the local safety. It first invokes a reachability graph constructor to build a one-step reachability graph $G_1$ to describe how the system evolves in one sampling step (Line 2). Then, a dynamic programming (DP) based approach finds the largest set $\Gamma_S = \{v_{s_1}, v_{s_2}, \ldots , v_{s_n}\}$ from which the system will not go out of the safe region. The $K$-step reachability graph $G_K$ is also built in the DP process based on $G_1$ (Line 3). After that, the tool searches the largest subset $\Gamma_I$ of $\Gamma_S$ that satisfies the inductiveness by using a reverse search algorithm (Line~4). The algorithm outputs $\Gamma_I$ as the target set $X_0$ (Line 5).

\begin{algorithm} [tbp]
    \SetAlgoLined
    \KwData{Dynamic system $f$ with safe state region $X$, the control law $\pi$, weakly-hard constraint $(m,K)$, sampling period $\delta$}
    \KwResult{Safe initial state set $X_0$}
    $\Gamma = \textbf{partition}(X,p)$\;
    \tcc{Search the grid set that satisfies local safety.}
    $G_1 = \textbf{constructOneStepGraph}()$ \;
    $\Gamma_S, G_K = \textbf{calculateLocalSafety}()$ \;
    \tcc{Search the grid set that satisfies inductiveness.}
    $\Gamma_I = \textbf{calculateInductivenessSet}()$ \;
    \Return{$\Gamma_I$}\;
\caption{Overall algorithm of SAW}
\label{algo:safe}
\end{algorithm}

The key functions of the tool are the reachability graph constructor, DP-based local safety set search, and reverse inductiveness set search. In the following sections, we introduce these three functions in detail.

\begin{algorithm} [tbp]
    \SetAlgoLined
    \KwData{Dynamic system $f$, grid set $\Gamma$, the control law $\pi$, sampling period $\delta$}
    \KwResult{Directed graph $G_1(\Gamma, E_1)$}
    \tcc{Initialize the edge set $E_1$ of $G_1$.}
    $E_1 \longleftarrow \emptyset$\;
    \For{$v \in \Gamma$}{
        \tcc{Consider deadline miss ($e=1$)/meet ($e=0$) respectively.}
        \For{$e \in \{0, 1\}$}{
            \tcc{Compute one step reachable set $R_1(v)$ from $v$.}
            $R_1(v) = \textbf{Flow*}(v, \delta, e)$\;
            \tcc{$v$ is unsafe and no edge is added if $X^c \cap R_1(v) \ne \emptyset$.}
            \lIf{$X^c \cap R_1(v) \ne \emptyset$}{\textbf{Conitnue}}
            \tcc{Add an edge pointing $v'$ from $v$ if $v' \cap R_1(v) \ne \emptyset$.}
            \For{$v' \in \Gamma$}{
                \lIf{$v' \cap R_1(v) \ne \emptyset$}{$E_1 \longleftarrow E_1 \cup \{(v, e, v')\}$}
            }
        }
    }
    \Return{$G_1(\Gamma, E_1)$}\;
\caption{Construct one-step graph: \textbf{constructOneStepGraph}()}
\label{algo:onestep}
\end{algorithm}

\subsection{Reachability Graph Construction}

Integration in dynamic system equations is often the most time-consuming part to trace the variation of the states. In this function, we use Flow* to get a valid overapproximation of reachable set (represented as flowpipes) starting from every grid after a sampling period $\delta$. Given a positive integer $n$, the graph constructed by the reachability set after $n$ sampling period, $n \cdot \delta$, is called a $n$-step graph $G_n$. Since the reachability for all the grids in any sampling step is independent under our grid assumption, we first build $G_1$ and then reuse $G_1$ to construct $G_K$ later without redundant computation of reachable set.

One-step graph is built with Algorithm~\ref{algo:onestep}. We consider deadline miss and deadline meet separately, corresponding to two categories of edges (Line 3). For a grid $v$, if the one-step reachable set $R_1(v)$ intersects with unsafe state $X^c$, then it is considered as an unsafe grid and we let its reachable grid be $\emptyset$. Otherwise, if $R_1(v)$ intersects with another grid $v'$ under the deadline miss/meet event $e$, then we add a directed edge $(v,e,v')$ from $v'$ to $v$ with label $e$. The number of outgoing edges for each grid node $v$ is bounded by $p^d$. Assuming that the complexity of Flow* to compute flowpipes for its internal clock $\epsilon$ is $O(1)$, we can get the overall time complexity as $O(|\Gamma| \cdot p^d \cdot \delta / \epsilon)$.


\begin{algorithm} [tbp]
    \SetAlgoLined
    \KwData{Directed graph $G_1(\Gamma, E_1)$, weakly-hard constraint $(m,K)$}
    \KwResult{Grid set $\Gamma_S$, directed graph $G_K(\Gamma, E_K)$}
    \For{$v \in \Gamma$}{
        \For{$n \leftarrow 0$ \KwTo $m$}{
            $\text{DP}(v, n, K) \longleftarrow \{v\}$\;
        }
    }
    \For{$k \leftarrow K - 1$ \KwTo $0$}{
        \For{$v \in \Gamma$}{
            \For{$n \leftarrow 0$ \KwTo $m$}{
                $\text{isSafe} \leftarrow True$\;
                \For{$e \in \{0, 1\}$}{
                    \If{$n + e \le m$} {
                        $\text{nextGrids}(v) \longleftarrow \{v' \mid (v, e, v') \in E_1\}$\; 
                        \lIf{$\text{nextGrids}(v) = \emptyset$}{$\text{isSafe} \leftarrow False$; \textbf{break}}
                        \For{$v' \in \text{nextGrids}(v)$}{
                            $R(v') \longleftarrow \text{DP}(v, n + e, k + 1)$\;
                            \lIf{$R(v) = \emptyset$}{$\text{isSafe} \leftarrow False$; \textbf{break}}
                            $\text{DP}(v', n, k) \longleftarrow \text{DP}(v', n, k) \cup R(v)$\;
                        }
                    }
                }
                \If{$\text{isSafe} = false$}{
                    $\text{DP}(v, n, k) \longleftarrow \emptyset$\;
                }
            }
        }
    }
    $\Gamma_S \longleftarrow \{v \mid \text{DP}(v, 0, 0) \ne \emptyset\}$\;
    $E_K \longleftarrow \{(v, v') \mid v \in \text{DP}(v, 0, 0)\}$\;
    \Return{$\Gamma_S, G_K(\Gamma, E_K)$}\;    
\caption{Search grid set for local safety: \textbf{calculateLocalSafety}()}
\label{algo:localsafe}
\end{algorithm}

$K$-step graph $G_K$ is built for finding the grid set that satisfies local safety and inductiveness. To avoid redundant computation on reachable set, we construct $G_K$ based on $G_1$ by traversing $K$-length paths, as the bi-product of local safety set searching procedure.


\subsection{DP-based Local Safety Set Search}
We propose a bottom-up dynamic programming for considering all the possible paths, utilizing the overlapping subproblems property (Algorithm~\ref{algo:localsafe}). The reachable grid set at step $K$ that is derived from a grid $v$ at step $k \leq K$ with respect to the number of deadline misses $n \leq m$ can be defined as $\text{DP}(v, n, k)$. To be consistent with Alogrithm \ref{algo:onestep}, this set is empty if and only if it does not satisfy the local safety. We need to derive $\text{DP}(v, 0, 0)$. Initially, the zero-step reachability is straight forward, i.e., $\forall u \in \Gamma, n \in [0, m]$, $\text{DP}(v, n, K) = \{v\}$. The transition is defined as:
\begin{displaymath}
    \forall k \in [0, K - 1]:\ \text{DP}(v, n, k) = \bigcup\limits_{\forall v', e: (v, e, v') \in E_1, n + e \le m} \text{DP}(v', n + e, k + 1).
\end{displaymath}
If there exists an empty set on the right hand side or there is no outgoing edge from $v$ for any $e$ such that $n + e \le m$, we let $\text{DP}(v, n, k) = \emptyset$. Finally, we have $\Gamma_S = \{v \mid \text{DP}(v, 0, 0) \neq \emptyset\}$, $E_K = \{(v, v') \mid v' \in \text{DP}(v, 0, 0)\}$.

We used $bitset$ to implement the set union which can accelerate 64 times under the 64-bit architecture. The time complexity is $O(|\Gamma|^2 / bits \cdot p^d \cdot K^2 + |\Gamma|^2)$, where $bits$ depends on the running environment. $|\Gamma|^2$ is contributed by $G_K$. 

\subsection{Reverse Inductiveness Set Search}

\begin{algorithm} [tbp]
    \SetAlgoLined
    \KwData{Directed graph $G_K(\Gamma, E_K)$, Grid set $\Gamma_S$}
    \KwResult{Grid set $\Gamma_I$}
    $\Gamma_U \longleftarrow \Gamma - \Gamma_S$\;
    \While{$\exists (v, v') \in E_K$ such that $v \notin \Gamma_U, v' \in \Gamma_U$}{
        $\Gamma_U \longleftarrow \Gamma_U \cup \{v\}$\;
    }
    $\Gamma_I = \Gamma - \Gamma_U$\;
    \Return{$\Gamma_I$}\;  
\caption{Search grid set for inductiveness: \textbf{calculateInductivenessSet}()}
\label{algo:inductiveness}
\end{algorithm}


To find the grid set $\Gamma_I \subseteq \Gamma_S$ that satisfies inductiveness, we propose a reverse search algorithm Algorithm~\ref{algo:inductiveness}. Basically, instead of directly searching $\Gamma_I$, we try to obtain $\Gamma_I$ by removing any grid $v$ within $\Gamma_S$, from which there exists a path reaching $\Gamma_U = \Gamma - \Gamma_S$. Specifically, Algorithm~\ref{algo:inductiveness} starts with initializing $\Gamma_U = \Gamma - \Gamma_S$ (line 1). The $\Gamma_U$ iteratively absorbs the grid $v$ that can reach $\Gamma_U$ in $K$ sampling periods, until a fixed point is reached (line 2-3). Finally $\Gamma_I = \Gamma - \Gamma_U$ is the largest set that satisfies inductiveness. It is implemented as a breadth first search (BFS) on the reversed graph of $G_K$, and the time complexity is $O(|\Gamma|^2)$.

%% file: sections/usage.tex
\section{Example Usage}\label{sec:usage}

\begin{example} \label{ex:1}
Consider the following linear control system from~\cite{Huang_HSCC19}:
\begin{displaymath}
    \begin{bmatrix} \dot{x_1} \\ \dot{x_2} \end{bmatrix}=\begin{bmatrix} 0 & 1 \\ 0 & -0.1 \end{bmatrix}\begin{bmatrix} x_1 \\ x_2 \end{bmatrix}+u,,\ \ \text{where}\ \
    u=\begin{bmatrix} 0 & 0 \\ -0.375 & -1.15 \end{bmatrix}\begin{bmatrix} x_1 \\ x_2 \end{bmatrix}.
\end{displaymath}
$\delta = 0.2$ and $step\_size = 0.01$. The initial state set is $x_1 \in [-1, 1]$ and $x_2 \in [-1, 1]$. The safe state set is $x_1 \in [-3, 3]$ and $x_2 \in [-3, 3]$. Following the input format shown in Listing~\ref{ls:format}. Thus, we prepare the model file as Listing~1.2.
\noindent\begin{minipage}{.47\textwidth}
\begin{lstlisting}[language=C++, caption=Input format, label = {ls:format}]
<state_dim> <input_dim> <grid_count>
<state_var_names> <input_var_names>
<state_ode.1>
...
<state_ode.state_dim>
<input_equa.1>
...
<input_equa.input_dim>
<period> <step_size>
<m> <k>
<safe_state.1>
...
<safe_state.state_dim>
<initial_state.1>
...
<initial_state.state_dim>
\end{lstlisting}
\end{minipage}\hfill
\begin{minipage}{.47\textwidth}
\begin{lstlisting}[language=C++, caption=example/model1.txt, label = ls:ex]
2 1 50
x1 x2 u
x2
-0.1 * x2 + u
-0.375 * x1 - 1.15 * x2
0.2 0.01
2 5
-3 3
-3 3
-1 1
-1 1
\end{lstlisting}
\end{minipage}
Then, we run our program with the model file.
\begin{lstlisting}[label = ls:ex1]
./saw example/model1.txt 
\end{lstlisting}
To further ease the use of our tool, we also pre-complied our tool for x86\_64 linux environment. In such environment, users do not need to compile our tool and can directly invoke \textbf{saw\_linux\_x86\_64} instead of \textbf{saw} (which is only available after manually compiling the tool).
\begin{lstlisting}[label = ls:ex2]
./saw_linux_x86_64 example/model1.txt
\end{lstlisting}

The program output is shown in Listing \ref{ls:result}. Line 6 shows the number of edges of $G_1$. Lines 8-10 provide the information of $G_K$, including the number of edges and nodes. Line 12 prints the safe initial set $X_0$. Our tool then determines whether the given initial set is safe by checking if it is the subset of $X_0$. 
\begin{lstlisting}[language=C++, caption=Verification result, label = ls:result]
[Info] Parsing model.
[Info] Building FLOW* configuration.
[Info] Building grids.
[Info] Building one-step graph.
       Process: 100.00%
[Success] Number of edges: 19354
[Info] Building K-step graph.
[Success] Start Region Size: 1908
          End Region: 1208
          Number of Edges: 102436
[Info] Finding the largest closed subgraph.
[Success] Safe Initial Region Size: 1622
[Info] Calculating area.
       Initial state region: 4.000000
       Grids Intersection:   4.000000
       Result: safe
\end{lstlisting}

\end{example}

%% file: sections/experiment.tex
\section{Experiments} \label{sec:experiment}

We implemented a prototype of SAW that is integrated with Flow*. In this section, we first compare our tool with the one-dimension abstraction~\cite{Huang_HSCC19}, on the full benchmarks from~\cite{Huang_HSCC19} (\#1--\#4) and also additional examples with no guarantee on exponential stability from related works (\#5 and \#6) \cite{prajna2004nonlinear}. Table~\ref{tab:benchmark} shows the benchmark settings, including the $(m, K)$ constraint set for each benchmark. Then, we show how different parameter settings affect the verification results of our tool. All our experiments were run on a desktop, with 6-core 3.60 GHz Intel Core i7.

\begin{table}[tbp] 
\caption{Benchmark setting. ODE denotes the ordinary differential equation of the example, $\pi$ denotes the control law, and $\delta$ is the discrete control stepsize.}
\small
\setlength\tabcolsep{3.3pt}
\renewcommand{\arraystretch}{1.2}
\begin{tabular}{|c|c|c|c|c|c|}\hline
    \# & ODE & $\pi$ & $\delta$ &  Safe State Set & $(m,K)$\\\hline
    \multirow{2}{*}{1}  & \multirow{2}{*}{\minitab[c]{$\dot{x}_1=x_2$\\$\dot{x}_2=-0.1x_2+u$}}
                        & \multirow{2}{*}{$u=-0.375x_1-1.15x_2$}
                        & \multirow{2}{*}{0.2}
                        & \multirow{2}{*}{\minitab[c]{$x_1{\in}[-3.0,3.0]$\\$x_2{\in}[-3.0,3.0]$}}
                        & \multirow{2}{*}{(2, 5)} \\ 
                        & & & & & \\\hline
    \multirow{2}{*}{2}  & \multirow{2}{*}{\minitab[c]{$\dot{x}_1=-2 x_1 + u_1$\\$\dot{x}_2=-0.9 x_2 + u_2$}}
                        & \multirow{2}{*}{\minitab[c]{$u_1=-x_1$ \\$u_2=-x_1 - x_2$}}
                        & \multirow{2}{*}{0.3}
                        & \multirow{2}{*}{\minitab[c]{$x_1{\in}[-6.0,6.0]$\\$x_2{\in}[-6.0,6.0]$}}
                        & \multirow{2}{*}{(1, 10)} \\ 
                        & & & & & \\\hline
    \multirow{2}{*}{3}  & \multirow{2}{*}{\minitab[c]{$\dot{x}_1=x_2 + u$\\$\dot{x}_2=-2x_1 - 0.1x_2  + u$}}
                        & \multirow{2}{*}{$u=x_1$}
                        & \multirow{2}{*}{1.6}
                        & \multirow{2}{*}{\minitab[c]{$x_1{\in}[-3.0,3.0]$\\$x_2{\in}[-3.0,3.0]$}}
                        & \multirow{2}{*}{(2, 10)} \\ 
                        & & & & & \\\hline
    \multirow{1}{*}{4}  & \multirow{1}{*}{$\dot{x}=x^2 - x^3 + u$}
                        & \multirow{1}{*}{$u=-2x$}
                        & \multirow{1}{*}{0.6}
                        & \multirow{1}{*}{$x{\in}[-4.0,4.0]$}
                        & \multirow{1}{*}{(2, 100)} \\\hline
    \multirow{1}{*}{5}  & \multirow{1}{*}{\minitab[c]{$\dot{x} = 0.2 x + 0.03 x^2 + u$}}
                        & \multirow{1}{*}{$u = -0.3 x^3$}
                        & \multirow{1}{*}{1.6}
                        & \multirow{1}{*}{\minitab[c]{$x \in [-2.0, 2.0]$}}
                        & \multirow{1}{*}{(1, 5)} \\\hline
    \multirow{2}{*}{6}  & \multirow{2}{*}{\minitab[c]{$\dot{x}_1=x_2 - x_1^3 + x_1^2$\\$\dot{x}_2=u$}}
                        & \multirow{2}{*}{\minitab[c]{$u=-1.22x_1 - 0.57x_2$ \\ $- 0.129x_2^3$}}
                        & \multirow{2}{*}{0.1}
                        & \multirow{2}{*}{\minitab[c]{$x_1{\in}[-5.0,5.0]$\\$x_2{\in}[-5.0,5.0]$}}
                        & \multirow{2}{*}{(2, 15)} \\
                        & & & & & \\\hline
\end{tabular}
\label{tab:benchmark}
\end{table}

\begin{table} [tbp] 
\caption{Experimental results. ExpParam denotes the parameters of the exponential stability, where ``N/A'' means that either the system is not exponentially stable or the parameters are not available. Initial state set denotes the set that needs to be verified. The last two columns denote the verification results of the one-dimension abstraction~\cite{Huang_HSCC19} and SAW, respectively. ``---'' means that no safe initial set $X_0$ is found by the tool. $p$ represents the partition number for each dimension in SAW. Time (in seconds) represents the execution time of SAW. }
\small
\setlength\tabcolsep{3.5pt}
\renewcommand{\arraystretch}{1.2} 
\begin{tabular}{|c|c|c|c|c|c|c|}
\hline
    \multirow{2}{*}{\#} & \multirow{2}{*}{ExpParam} & \multirow{2}{*}{Initial State Set} & One-dimension Abstraction & \multicolumn{3}{c|}{SAW} \\ \cline{4-7}
     &  &  & Result & p & Result & Time \\\hline
    \multirow{2}{*}{1}  & \multirow{2}{*}{\minitab[c]{$\alpha=1.8$,\\ $\lambda=0.4$.}}
                        & \multirow{2}{*}{\minitab[c]{$x_1{\in}[-1.0,1.0]$ \\ $x_2{\in}[-1.0,1.0]$}}
                        & \multirow{2}{*}{---} 
                        & \multirow{2}{*}{50}
                        & \multirow{2}{*}{Yes} 
                        & \multirow{2}{*}{72.913}\\ 
                        & & & & & & \\\hline
    \multirow{2}{*}{2}  & \multirow{2}{*}{\minitab[c]{$\alpha=1.1$,\\ $\lambda=1.8$.}}
                        & \multirow{2}{*}{\minitab[c]{$x_1{\in}[-6.0,6.0]$ \\ $x_2{\in}[-6.0,6.0]$}}
                        & \multirow{2}{*}{\minitab[c]{No \\($X_0: x^2_1 + x^2_2 \leq 1.947^2$)}}
                        & \multirow{2}{*}{30} 
                        & \multirow{2}{*}{Yes} 
                        & \multirow{2}{*}{10.360}\\ 
                        & & & & & & \\\hline
    \multirow{2}{*}{3}  & \multirow{2}{*}{\minitab[c]{$\alpha=2$,\\ $\lambda=0.37$.}}
                        & \multirow{2}{*}{\minitab[c]{$x_1{\in}[-1.0,2.0]$ \\ $x_2{\in}[-1.0,1.0]$}}
                        & \multirow{2}{*}{---}
                        & \multirow{2}{*}{100} 
                        & \multirow{2}{*}{Yes}
                        & \multirow{2}{*}{183.30}\\ 
                        & & & & & & \\\hline
    \multirow{2}{*}{4}  & \multirow{2}{*}{\minitab[c]{$\alpha=1.4$,\\ $\lambda=1$.}}
                        & \multirow{2}{*}{$x{\in}[-4.0,4.0]$}
                        & \multirow{2}{*}{---}
                        & \multirow{2}{*}{30} 
                        & \multirow{2}{*}{Yes}
                        & \multirow{2}{*}{80.613}\\ 
                        & & & & & & \\\hline
    \multirow{1}{*}{5}  & \multirow{1}{*}{N/A}
                        & \multirow{1}{*}{\minitab[c]{$x{\in}[-1.56,1.32]$}}
                        & \multirow{1}{*}{---}
                        & \multirow{1}{*}{100} 
                        & \multirow{1}{*}{Yes}
                        & \multirow{1}{*}{4.713}\\\hline
    \multirow{2}{*}{6}  & \multirow{2}{*}{N/A}
                        & \multirow{2}{*}{\minitab[c]{$x_1{\in}[-5.0,5.0]$ \\ $x_2{\in}[-5.0,5.0]$}}
                        & \multirow{2}{*}{---}
                        & \multirow{2}{*}{50} 
                        & \multirow{2}{*}{Yes}
                        & \multirow{2}{*}{750.77}\\ 
                        & & & & & & \\\hline
\end{tabular}
\label{tab:results}
\end{table}

\subsection{Comparison with One-dimension Abstraction}

Table~\ref{tab:results} shows the experimental results. It is worth noting that the one-dimension abstraction cannot find the safe initial set in most cases from~\cite{Huang_HSCC19}. In fact, it only works effectively for a limited set of $(m,K)$, e.g., when no consecutive deadline misses is allowed. For general $(m,K)$ constraints, one-dimension abstraction performs much worse due to the over-conservation. Furthermore, we can see that, without exponential stability, one-dimension abstraction based approach is not applicable for the benchmarks \#5 and \#6. Note that for benchmark \#2, one-dimension abstraction obtains a non-empty safe initial set $X_0$, which however, does not contain the given initial state set. Thus we use ``No'' instead of ``---'' to represent this result. Conversely, for every example, our tool computes a feasible $X_0$ that contains the initial state set (showing the initial state set is safe), which we denote as ``Yes''.


\subsection{Impact of $(m,K)$, Granularity, and Stepsize}

\begin{figure}[tbp]
\centering
	\subfloat[][$(m, K) = (2, 5)$]{%
		\includegraphics[width=0.33\textwidth]{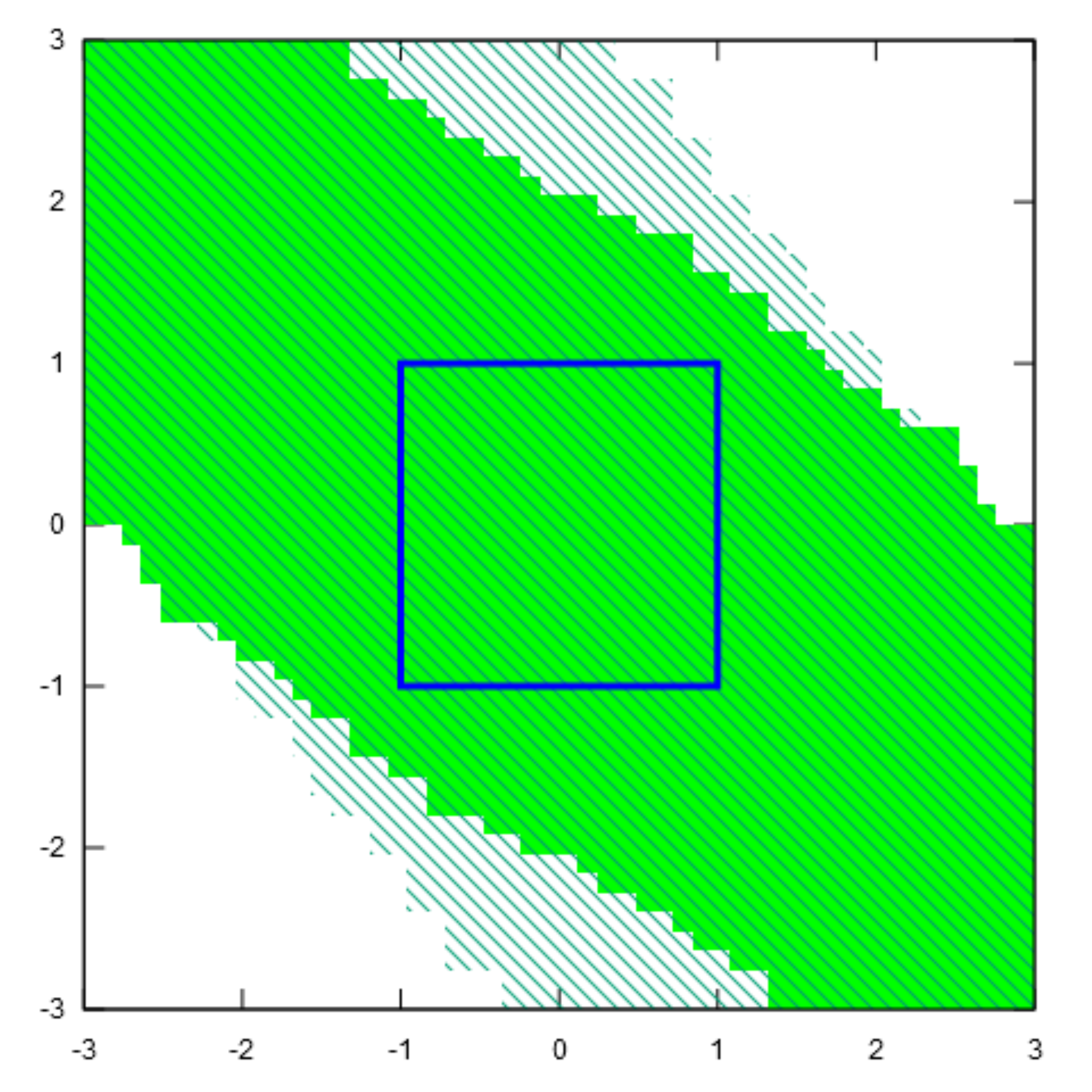}%
		\label{fig:mk1}%
	}
	\subfloat[][$(m, K) = (2, 9)$]{%
		\includegraphics[width=0.33\textwidth]{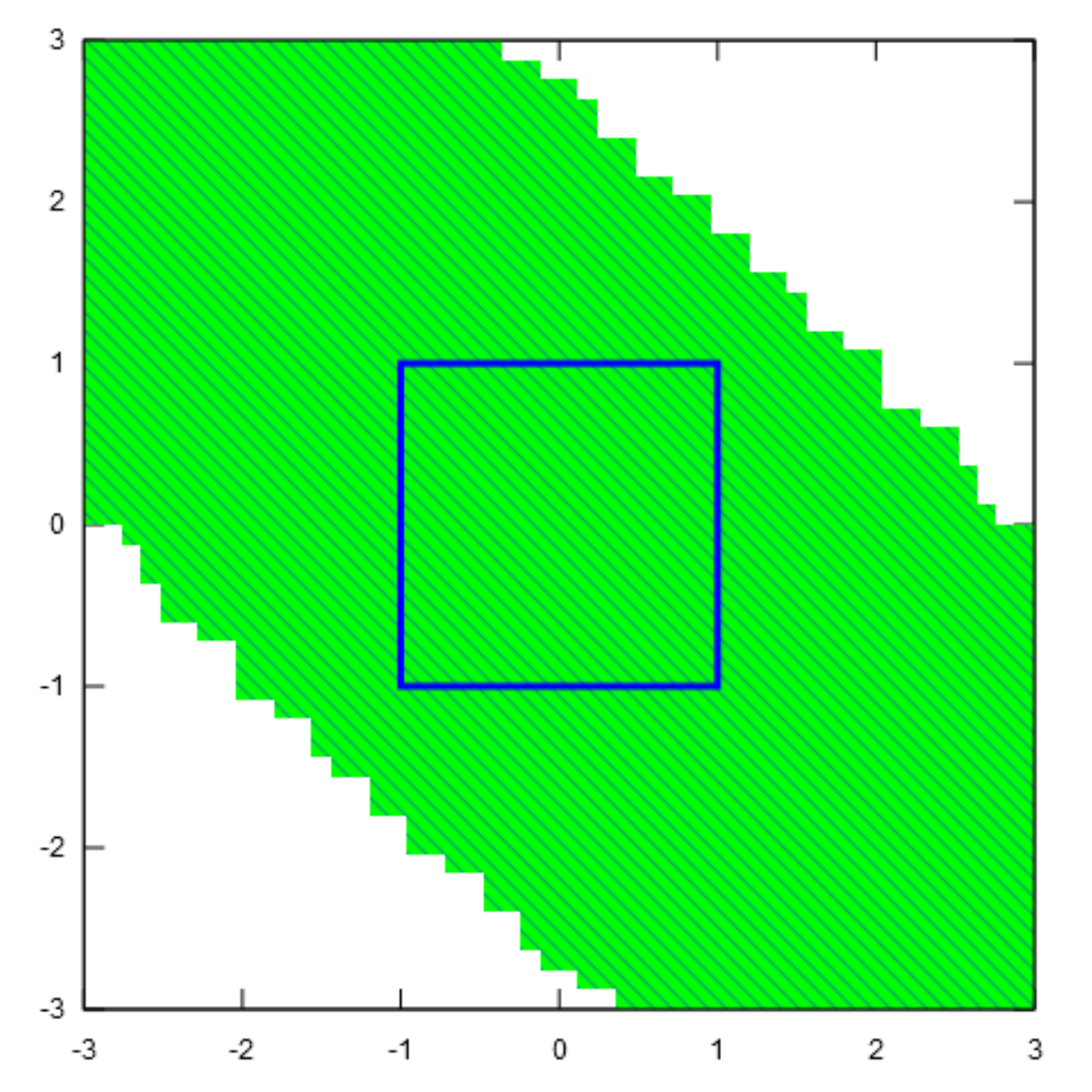}%
		\label{fig:mk3}%
	}
	\subfloat[][$(m, K) = (3, 9)$]{%
		\includegraphics[width=0.33\textwidth]{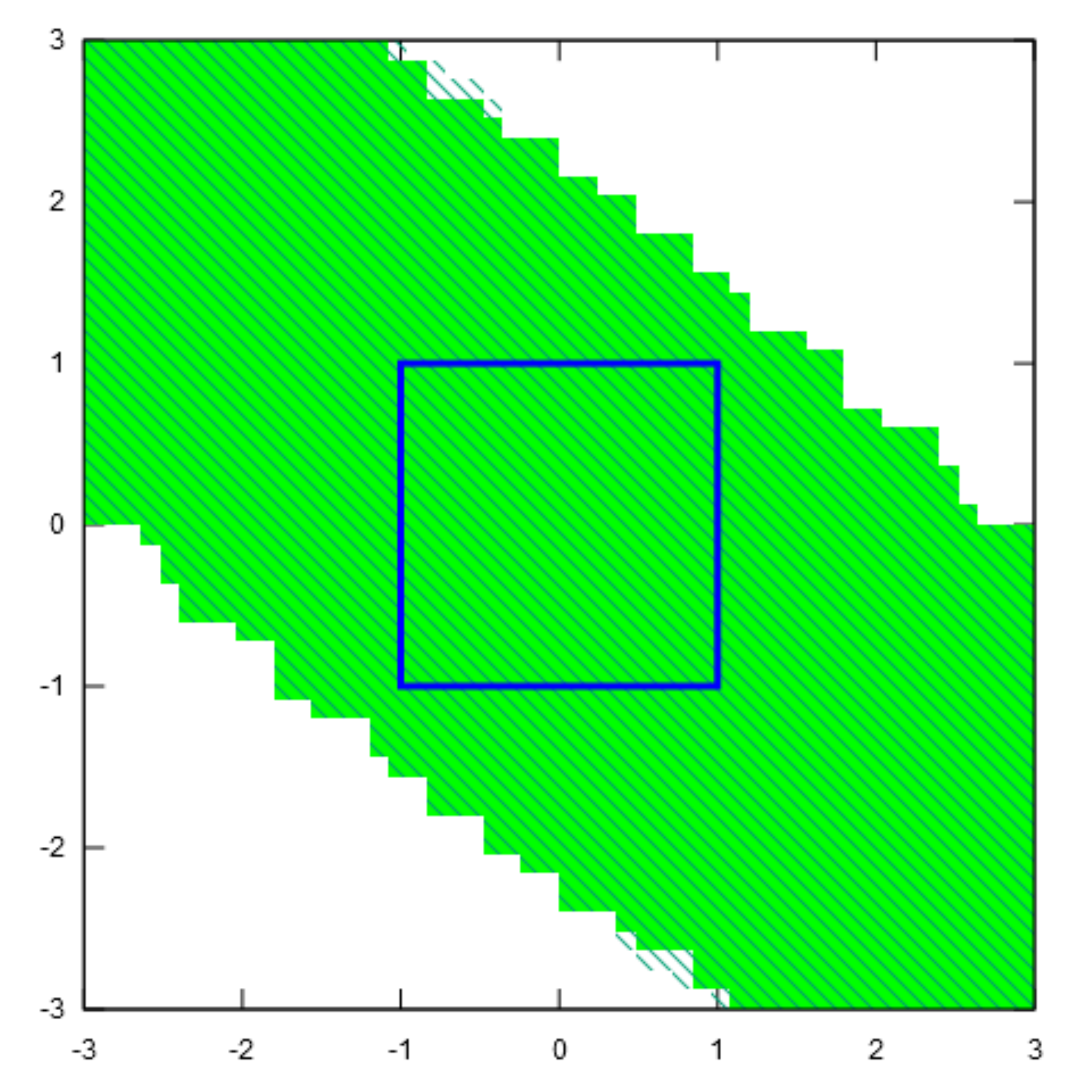}%
		\label{fig:mk4}%
	} \\
	\subfloat[][$p = 15$]{%
		\includegraphics[width=0.33\textwidth]{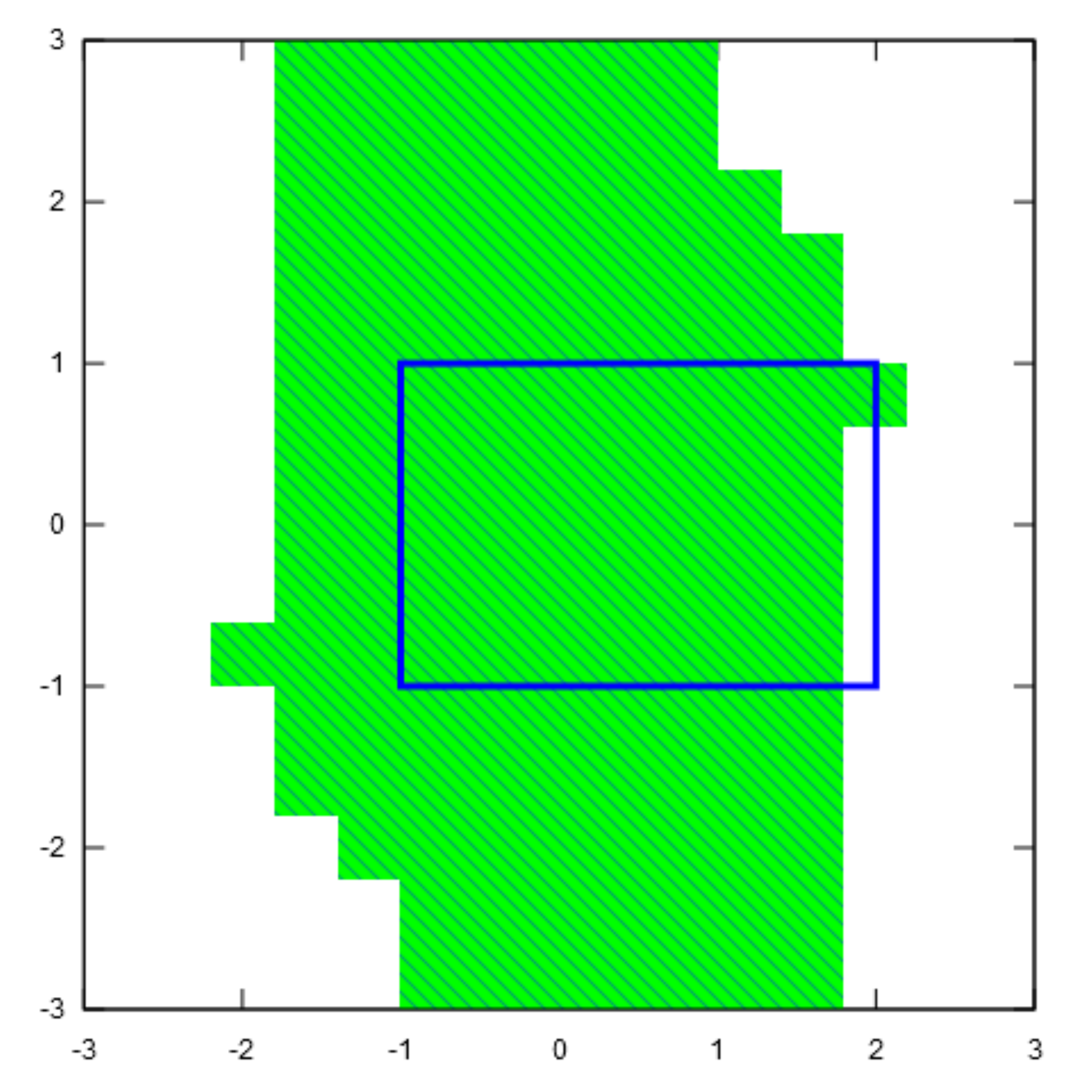}%
		\label{fig:p1}%
	}
	\subfloat[][$p=20$]{%
		\includegraphics[width=0.33\textwidth]{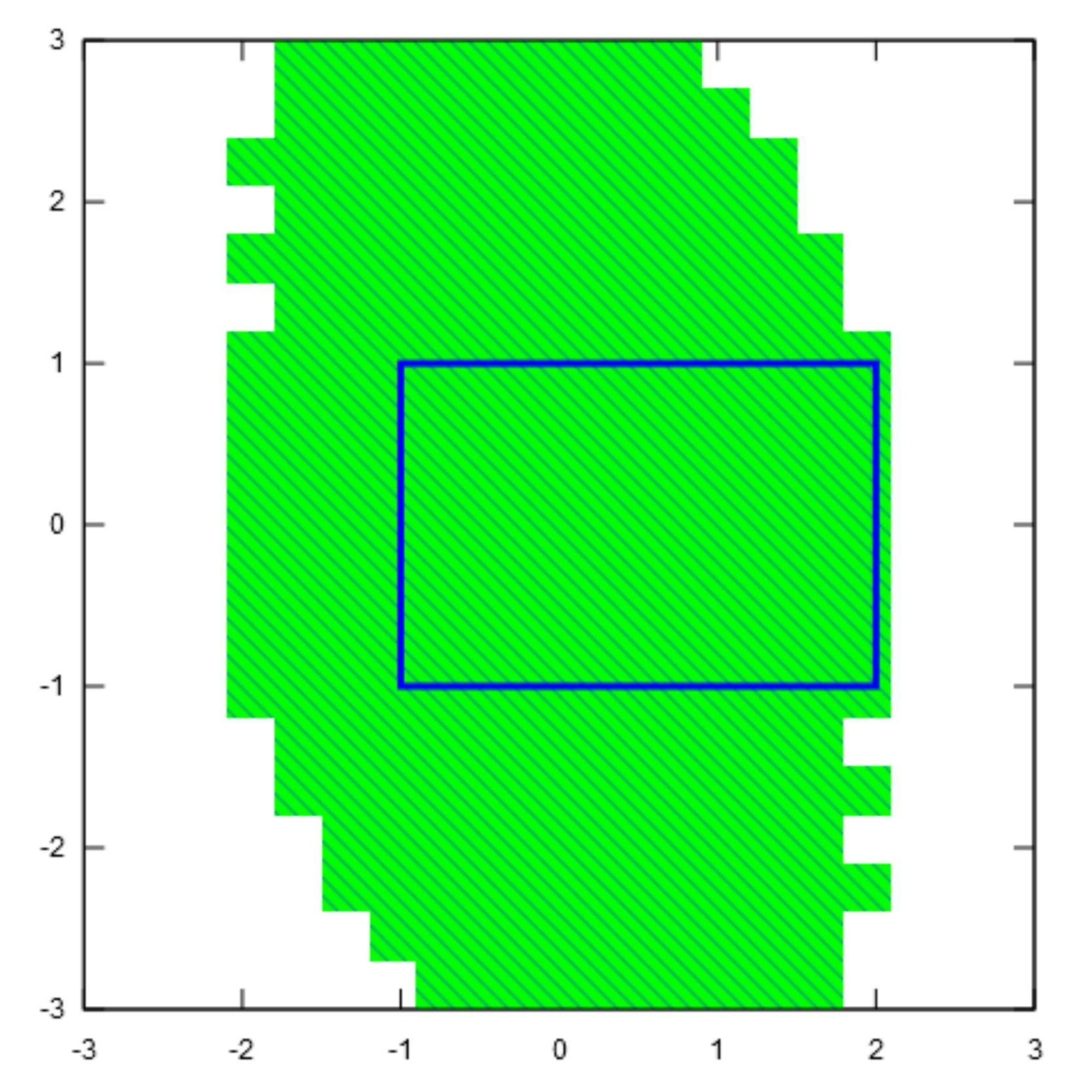}%
		\label{fig:p2}%
	}
	\subfloat[][$p=100$]{%
		\includegraphics[width=0.33\textwidth]{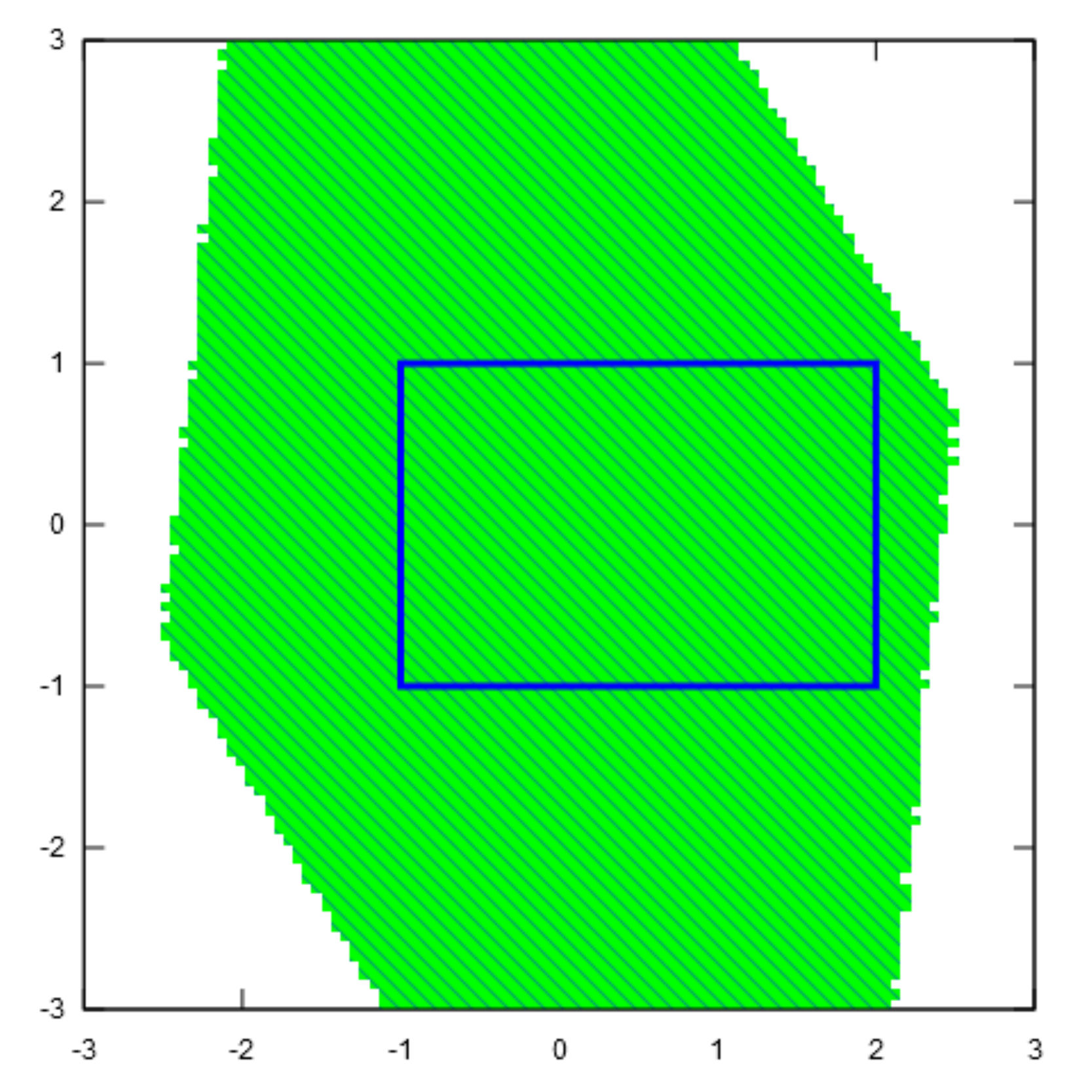}%
		\label{fig:p3}%
	}
    \caption{Results under different $(m, K)$ values (Figures~\ref{fig:mk1}, \ref{fig:mk3}, \ref{fig:mk4}) and different granularities (Figures~\ref{fig:p1}, \ref{fig:p2}, \ref{fig:p3}). The green solid region is $\Gamma_I$. The slashed region is $\Gamma_S$. The blue rectangle is the initial state set that needs to be verified.}
    \label{fig:example}
\end{figure}

\noindent\textbf{$(m,K)$}. We take benchmark \#1 (Example~\ref{ex:1} in Section~\ref{sec:usage}) as an example and run our tool under different $(m, K)$ values. Figures~\ref{fig:mk1}, \ref{fig:mk3}, \ref{fig:mk4} demonstrate that, for this example, the size of local safety region $\Gamma_S$ shrinks when $K$ gets larger. The size of inductiveness region $\Gamma_I$ grows in contrast. $\Gamma_S$ becomes the same as $\Gamma_I$ when $K$ gets larger, in which case $m$ is the primary parameter that influences the size of $\Gamma_I$.

\smallskip
\noindent\textbf{Granularity.} We take benchmark \#3 as an example, and run our tool with different partition granularities. The results (Figures~\ref{fig:p1}, \ref{fig:p2}, \ref{fig:p3}) show that $\Gamma_I$ grows when $p$ gets larger. The choice of $p$ has significant impact on the result (e.g., the user-defined initial state set cannot be verified when $p = 15$).

\smallskip
\noindent\textbf{Stepsize.} We take benchmark \#5 as an example, and run our tool with different stepsizes of Flow*. With the same granularity $p = 100$, we get the safe initial state set $\Gamma_I = [-1.56, 1.32]$ when $step\_size = 0.1$, but $\Gamma_I$ is empty when $step\_size = 0.3$. The computation times are 4.713 sec and 1.835 sec, respectively. Thus, we can see that there is a trade-off between the computational efficiency and the accuracy. 

%% file: sections/conclusion.tex
\section{Conclusion} \label{sec:conclusion}

In this paper, we present a new tool SAW to compute a tight estimation of safe initial set for infinite-time safety verification of general nonlinear weakly-hard systems. The tool first discretizes the safe state set into grids. By constructing a reachability graph for the grids based on existing tools, the tool leverages graph theory and dynamic programming technique to compute the safe initial set. We demonstrate that our tool can significantly outperform the state-of-the-art one-dimension abstraction approach, and analyze how different constraints and parameters may affect the results of our tool. 
Future work includes further speedup of the reachability graph construction via parallel computing.